\documentclass[prc,twocolumn,superscriptaddress,showpacs,amssymb,amsmath,amsfonts,aps]{revtex4}
\usepackage{graphicx}
\usepackage{dcolumn}
\usepackage{epsfig}
\begin{document}
\title{Nucleon electromagnetic form factors
and electroexcitation of low lying nucleon resonances 
in a light-front relativistic quark model \\}

\newcommand*{\JLAB }{ Thomas Jefferson National Accelerator Facility, 
Newport News, Virginia 23606, USA}
\affiliation{\JLAB }
\newcommand*{\YEREVAN }{ Yerevan Physics Institute, 375036 Yerevan, 
Armenia}
\affiliation{\YEREVAN }
\author{I.G.~Aznauryan}
     \affiliation{\JLAB}
     \affiliation{\YEREVAN}
\author{V.D.~Burkert}
     \affiliation{\JLAB}
\begin{abstract}
{
We utilize a light-front relativistic quark model (LF RQM)
to predict the $3q$ core contribution to the
electroexcitation amplitudes
for the $\Delta(1232)$P$_{33}$,  N(1440)P$_{11}$,
N(1520)D$_{13}$, and N(1535)S$_{11}$
up to $Q^2= 12~$GeV$^2$. The parameters of the model
have been specified via description of the nucleon electromagnetic
form factors in the approach that combines
$3q$ and pion-cloud contributions
in the LF dynamics.
}
\end{abstract}
\maketitle

\section{Introduction}
In the past decade, with the advent of the
new generation of electron beam facilities,
there has been dramatic progress in studies of the
electroexcitation of nucleon resonances that resulted in 
more reliable extractions of the resonance electrocouplings, and 
a significant extension of the $Q^2$ range.
The most accurate and complete information has been obtained
for  the four lowest excited states, 
which have been measured in a range of $Q^2$
up to $8$~GeV$^2$ for the $\Delta(1232)P_{33}$ and $N(1535)S_{11}$, 
and up to $4.5~$GeV$^2$ for the $N(1440)P_{11}$ and $N(1520)D_{13}$ 
(see reviews \cite{Aznauryan11,Tiator11}).

At relatively small $Q^2$, nearly massless pions generate
pion-loop contributions that may significantly alter 
quark model predictions. It is expected
that the corresponding hadronic component, including
contributions from other mesons, will be rapidly losing 
strength with increasing
$Q^2$. The Jefferson Lab 12 GeV upgrade
will open up a new era in the exploration of excited
nucleons when the quark core of the nucleon  
and its excited states will be more 
fully exposed to the electromagnetic probe.

The aim of this paper is to estimate the $3q$ core contribution
to the electrocoupling amplitudes of the $\Delta(1232)P_{33}$, 
$N(1440)P_{11}$, $N(1520)D_{13}$, and  $N(1535)S_{11}$. 
The approach we use is based on the LF
dynamics which presents the most suitable framework
for describing the transitions between relativistic
bound systems \cite{Drell,Terentiev,Brodsky}.
In early works by Berestetsky and Terent'ev \cite{Terentiev}, 
the approach was based on the construction
of the generators of the Poincar\'e group in the LF.
It was later formulated 
in the infinite momentum frame (IMF) \cite{Terentiev1,Aznauryan1}.
This allowed one to demonstrate more clearly 
that diagrams which violate 
impulse approximation, i.e. the diagrams
containing  vertices like 
$\gamma^*\rightarrow q{\bar q}$, do not contribute.  
The interpretation of results for 
$\gamma^* N\rightarrow N(N^*)$ in terms 
of the vertices
$N(N^*)\leftrightarrow 3q$ 
and corresponding wave functions became more evident.  
In Refs. \cite{Aznauryan1,Aznauryan2,Aznauryan85,Aznauryan3}, 
the LF RQM formulated in IMF
was utilized for the investigation of 
nucleon form factors and the electroexcitation
of nucleon resonances. 
These observables were investigated also
in the LF Hamiltonian dynamics in Ref. \cite{Capstick1}.
In both cases
a complete orthogonal set of
wave functions has been used, 
that correspond to the classification
of the nucleon and nucleon resonances
within the group $SU(6)\times O(3)$;
the relativistic-covariant form of these wave functions
has been found in Ref. \cite{Aznauryan1}.
We specify the parameters of the model for the $3q$ contribution
via description of the electromagnetic form factors, 
by combining the $3q$ and pion-cloud contributions.
The pion-cloud contribution
has been incorporated using the LF approach of Ref. \cite{Miller}.

In Sec. II we present briefly the formalism to compute
the $3q$ contribution to the $\gamma^* N\rightarrow N(N^*)$ amplitudes.
In Sec. III we discuss the description of nucleon electromagnetic
form factors
at $0\leq Q^2< 16~$GeV$^2$. To achieve 
description of experimental data at $Q^2>0$,
we incorporate the $Q^2$-dependence of the constituent quark mass that is 
expected from the lattice QCD and Dyson-Schwinger equations approach
\cite{Bowman,Bhagwat1,Bhagwat2}.
With the LF RQM specified via description of the nucleon 
electromagnetic form factors, we predict in Sec. IV
the quark core contribution to the electroexcitation
amplitudes of the aforementioned resonances at $Q^2 \le 12~$GeV$^2$.
The results are summarized in Sec. V.  

\section{Quark core contribution to transition amplitudes}
The $3q$ contribution to the $\gamma^* N\rightarrow N(N^*)$
transitions has been evaluated within the approach
of Refs. \cite{Aznauryan1,Aznauryan2} where the 
LF RQM
is formulated in the IMF.
The IMF is chosen in such a way, that the 
initial hadron moves
along the $z$-axis with the momentum $P_z\rightarrow \infty$,
the virtual photon momentum is
$ k^{\mu}=\left(
\frac {m_{out}^2-m_{in}^2-\mathbf{Q}^2_{\perp}}{4P_z},
\mathbf{Q}_{\perp}, 
-\frac {m_{out}^2-m_{in}^2-\mathbf{Q}^2_{\perp}}{4P_z}\right)$,
the final hadron momentum is
$P'=P+k$, and $Q^2\equiv -k^2=\mathbf{Q}_{\perp}^2$; $m_{in}$ and $m_{out}$
are masses of the initial and final hadrons, respectively. 
The matrix elements of the electromagnetic current
are related to the $3q$-wave functions in the following way:
\begin{eqnarray}
&& \frac{1}{2P_z}<N(N^*),S'_z|J_{em}^{0,3}|N,S_z>|_
{P_z\rightarrow\infty} \nonumber \\
&&=3e\int \Psi'^+(p'_a,p'_b,p'_c) Q_a
\Psi(p_a,p_b,p_c) d\Gamma,
\label{eq:sec1}
\end{eqnarray}
where $S_z$ and $S'_z$ are the projections of the hadron
spins on the $z$-direction.
In Eq. (\ref{eq:sec1}), it is supposed that 
the photon interacts with quark $a$ (the quarks
in hadrons are denoted by $a,b,c$), 
$Q_a$ 
is the charge of this quark in units of $e$ ($e^2/4\pi=1/137$);
$\Psi$ and $\Psi'$ are wave functions
in the vertices $N(N^*)\leftrightarrow 3q$;
$p_i$ and $p'_i$ ($i=a,b,c$) are the quark momenta;
$d\Gamma$ is the phase space volume.
The relations between the matrix elements (\ref{eq:sec1})
and the $\gamma^* N\rightarrow N(N^*)$ form factors and transition 
helicity amplitudes are given in Appendix.

We parametrize the quark momenta in the IMF via:
\begin{eqnarray}
&&\mathbf{p}_i=x_i\mathbf{P}+\mathbf{q}_{i\perp},~
~\mathbf{p}'_i=x_i\mathbf{P}'+\mathbf{q}'_{i\perp},
~~\sum\limits_{i} {x_i}=1,
\label{eq:sec2}\\
&&\mathbf{P}\mathbf{q}_{i\perp}=\mathbf{P}'\mathbf{q}'_{i\perp}=0,~~~
\sum\limits_{i} {\mathbf{q}_{i\perp}}
=\sum\limits_{i} {\mathbf{q}'_{i\perp}}=0,
\label{eq:sec3}
\end{eqnarray}
where 
$\mathbf{q}'_{i\perp}=\mathbf{q}_{i\perp}-y_i\mathbf{Q}_{\perp}$
and $y_a=x_a-1$, $y_b=x_b$, $y_c=x_c$.
The phase space volume corresponding to this parametrization is:
\begin{equation}
d\Gamma=(2\pi)^{-6}\frac
{d\mathbf{q}_{b\perp}d\mathbf{q}_{c\perp}dx_b dx_c}
{4x_ax_bx_c}.
\label{eq:sec4}
\end{equation}

According to results of Ref. \cite{Aznauryan1},
obtained through relativistic-covariant transformation,
the wave function $\Psi$ is related
to the wave function in the c.m.s. of the 
system of three quarks 
through Melosh matrices \cite{Melosh}:
\begin{equation}
\Psi=U^+(p_a)U^+(p_b)U^+(p_c)\Psi_{fss}
\Phi(\mathbf{q}_{a},\mathbf{q}_{b},\mathbf{q}_{c}).
\label{eq:sec5}
\end{equation}
Here we have separated the flavor-spin-space ($\Psi_{fss}$)
and spatial ($\Phi$)
parts of the wave function 
of the quarks in their c.m.s.
The Melosh matrices are defined by
\begin{equation}
U(p_i)=\frac{m_q+M_0x_i+i\epsilon _{lm}\sigma_l q_{im}}
{\sqrt{(m_q+M_0x_i)^2+\mathbf{q}_{i\perp}^2}},
\label{eq:sec6}
\end{equation}
where $m_q$ is the quark mass and
$M_0$ is invariant mass
of the system of initial quarks:
\begin{equation}
M_0^2=\left(\sum\limits_{i} {p_i}\right)^2=
\sum\limits_{i} {\frac{\mathbf{q}_{i\perp}^2+m_q^2}{x_i}}.
\label{eq:sec7}
\end{equation}
In the 3-quark c.m.s.:
\begin{equation}
M_0^2=\left(\sum\limits_{i} 
{\omega_i}\right)^2,~~~\omega_i=\sqrt{m_q^2+\mathbf{q}_{i}^2},
~~~q_{iz}+\omega_i=M_0x_i.
\label{eq:sec8}
\end{equation}
We construct the flavor-spin-space
parts of the wave functions 
by utilizing
the rules \cite{Capstick1,Isgur1}
that correspond to the classification
of the nucleon and nucleon resonances 
within the group $SU(6)\times O(3)$.

For the final state quarks, 
the quantities defined in Eqs. (\ref{eq:sec5}-\ref{eq:sec8}) 
are expressed through $p'_i$, $\mathbf{q}'_{i}$,
and $M'_0$.

To study sensitivity to the form of the quark wave function,
we employ two forms of the spatial wave function:

\begin{eqnarray} 
&&\Phi_1\sim exp(-M_0^2/6\alpha_1^2),
\label{eq:sec9}
\\
&&\Phi_2\sim  
exp\left[-({\bf{q}}_1^2+{\bf{q}}_2^2+{\bf{q}}_3^2)/2\alpha_2^2\right],
\label{eq:sec10}
\end{eqnarray}
that were used, respectively, in Refs. 
\cite{Terentiev,Aznauryan1,Aznauryan2}
and \cite{Capstick1}. 
 
\section{Nucleon}

The nucleon 
electromagnetic form factors were described by
combining the
$3q$ and $\pi N$ contributions
to the nucleon wave function.
With the pion loops evaluated according to
Ref. \cite{Miller},   
the nucleon wave function has the form: 
\begin{equation}
|N>=0.95|3q>+0.313|\pi N>,
\label{eq:nuc1}
\end{equation}
where the portions of different contributions 
were found from the condition the charge
of the proton be unity: $F_{1p}(0)=1$.

The values of the quark mass $m_q$ and of the parameters
$\alpha_{1,2}$  for the
 wave functions (\ref{eq:sec9},\ref{eq:sec10})
were found from the description
of $\mu_p=G_{Mp}(0)$ and $\mu_n=G_{Mn}(0)$.
Best results,
\begin{equation}
\mu_p=2.86\frac{e}{2m_N},~~~
\mu_n=-1.86\frac{e}{2m_N},
\label{eq:nuc2}
\end{equation}
were obtained
with $m_q(0)=0.22~$GeV and
\begin{equation}
\alpha_1=0.37~{\rm GeV},~~~\alpha_2=0.41~{\rm GeV}.
\label{eq:nuc3}
\end{equation}
The quark mass $m_q(0)=0.22~$GeV 
coincides with the value obtained  
from the description of the spectrum
of baryons and mesons and their excited states
in the relativized quark model \cite{Isgur2,Capstick2}.

The parameters (\ref{eq:nuc3}), that correspond 
to different forms of wave functions (\ref{eq:sec9})
and (\ref{eq:sec10}),
give very close magnitudes for
the mean values of invariant masses and momenta
of quarks at $Q^2=0$:
$<M^2_0>\approx 1.35~$GeV$^2$ and 
$<{\mathbf{q}}_{i}^2>\approx 0.1~$GeV$^2,~i=a,b.c$.
  
\begin{figure*}[htp]
\begin{center}
\includegraphics[width=12.0cm]{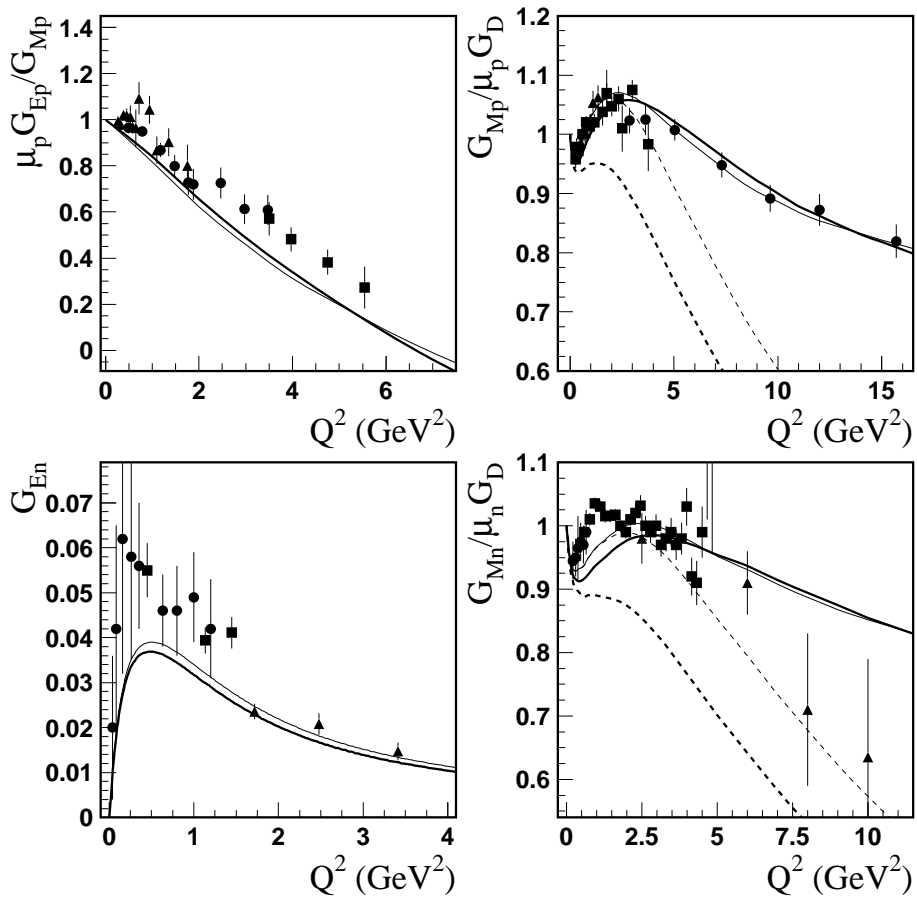}
\vspace{-0.1cm}
\caption{\small
Nucleon electromagnetic form factors.
The curves present the results obtained
taking into account two contributions to the nucleon
(Eq. \ref{eq:nuc1}): the pion-cloud  
and the $3q$ core.
The thick and thin curves
correspond, respectively, to 
the wave functions (\ref{eq:sec9}) and 
(\ref{eq:sec10}).
The solid curves are the results obtained 
 with the running quark masses 
(\ref{eq:nuc4},\ref{eq:nuc5}) and
the dashed  curves 
correspond to
the constant quark mass.
Data for $G_{Ep}(Q^2)/G_{Mp}(Q^2)$ are from Refs.
\cite{Jones} - circles, \cite{Gayou} - boxes,
\cite{Price} - triangles;
for $G_{Mp}(Q^2)$: 
\cite{Sill} - circles, \cite{Bartel} - boxes,
\cite{Price} - triangles;
for $G_{En}(Q^2)$ :
\cite{Sick} - circles, \cite{Madey} - boxes,
\cite{Riordan} - triangles;
for $G_{Mn}(Q^2)$:
\cite{Anderson} - circles, \cite{Lachniet} - boxes,
\cite{Rock} - triangles.
\label{fig:nucleon}}
\end{center}
\end{figure*}

A constant value of the quark mass 
gives rise to rapidly decreasing  
form factors; for  $G_{Mp}(Q^2)$ and $G_{Mn}(Q^2)$
this is demonstrated in Fig. \ref{fig:nucleon}. 
The wave functions (\ref{eq:sec9},\ref{eq:sec10}) 
increase as $m_q$ decreases. Therefore, to describe the
experimental data we have assumed the $Q^2$-dependent
constituent quark mass. We have used two different 
parameterizations of this mass:
\begin{eqnarray}
&& m_q^{(1)}(Q^2)=\frac{0.22{\rm GeV}}{1+Q^2/56{\rm GeV}^2},
\label{eq:nuc4}\\
&& m_q^{(2)}(Q^2)=\frac{0.22{\rm GeV}}{1+Q^2/18{\rm GeV}^2}
\label{eq:nuc5}
\end{eqnarray}
for the wave functions $\Phi_1$ and $\Phi_2$, respectively. This
resulted in a good description of the nucleon electromagnetic 
form factors for $Q^2\le 16~$GeV$^2$.

\begin{figure*}[htp]
\begin{center}
\includegraphics[width=12.0cm]{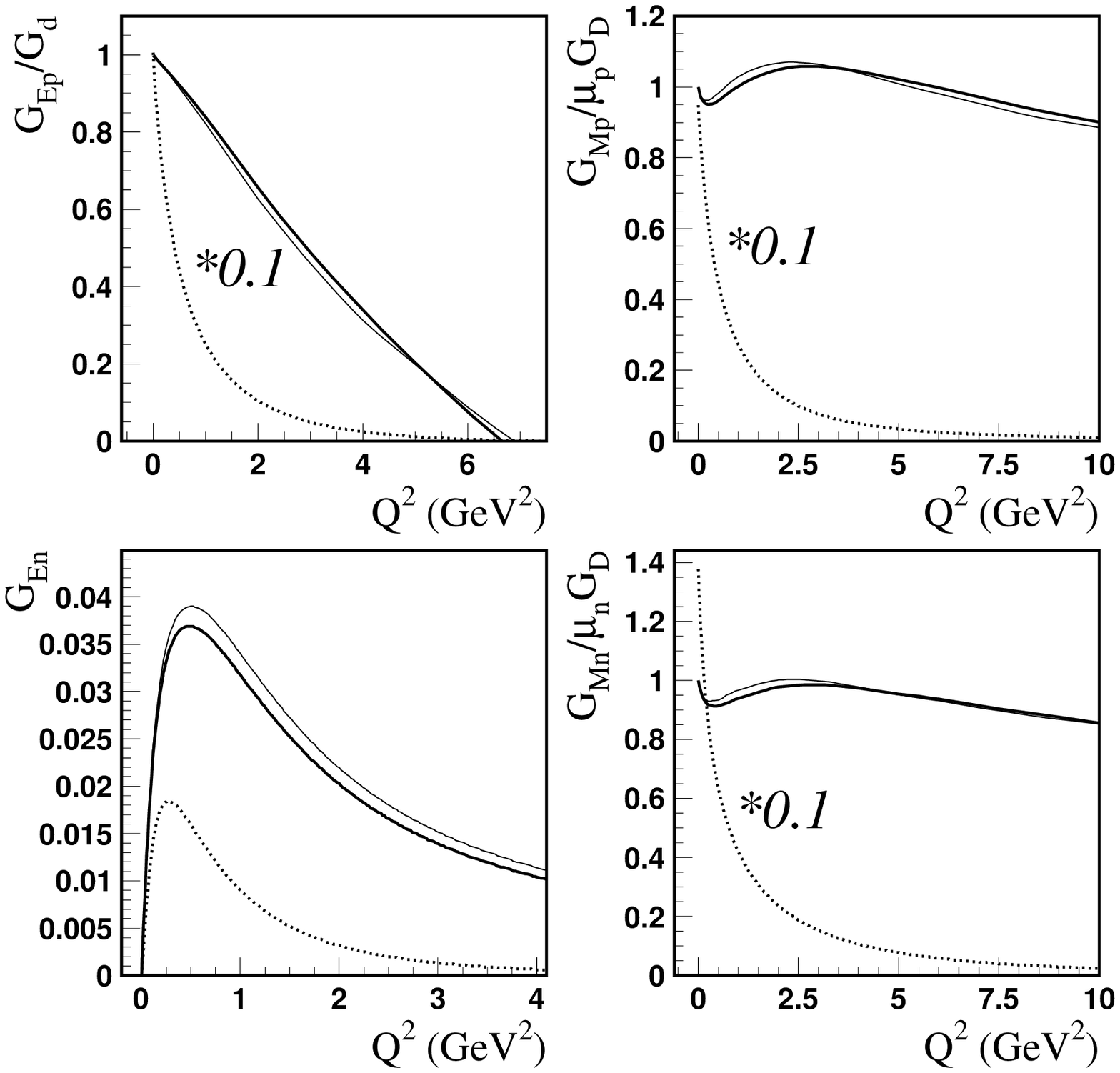}
\vspace{-0.1cm}
\caption{\small
Nucleon electromagnetic form factors.
The legend for
the solid  curves
is as for Fig. \ref{fig:nucleon}.
The dotted curves are the pion-cloud
contributions \cite{Miller}; for all form factors,
except $G_{En}(Q^2)$, the shown results 
for these contributions 
should be multiplied by 0.1.
\label{fig:cloud}}
\end{center}
\end{figure*}

In Fig.~\ref{fig:cloud} we show separately 
the pion-cloud contributions. Clearly, at $Q^2>2~$GeV$^2$,
all form factors are dominated by the $3q$-core contribution.                                     

The $Q^2$-dependence of the constituent quark mass
(Eqs. \ref{eq:nuc4},\ref{eq:nuc5}) 
is in qualitative agreement with the QCD lattice calculations
and Dyson-Schwinger equations
\cite{Bowman,Bhagwat1,Bhagwat2}, where
the running quark mass is generated dynamically.
However, we want to point out that there is no direct connection between
the functional forms of these masses. 
In QCD lattice calculations
and Dyson-Schwinger equations
we deal with quarks that
do not possess a mass-shell,
and the running quark mass is a function
of its virtuality, i.e. the quark four-momentum square.
In constituent quark models,
including the LF approaches 
\cite{Terentiev,Aznauryan1,Aznauryan2,Capstick1},
the quarks are mass-shell objects
(see Eqs. \ref{eq:sec7},\ref{eq:sec8}).
In LF RQM, the virtuality of quarks is characterized by invariant masses 
of the 3-quark system:
$M_0^2$ and ${M'}_0^2$.
Mean values of $M_0^2$ and ${M'}_0^2$ are equal to each other  
and are increasing with increasing 
$Q^2$.

The mechanism, that generates the running quark mass,
can generate also quark
anomalous magnetic moments and
form factors \cite{Roberts}. 
In our approach, we have obtained good description
of the nucleon electromagnetic form factors
without introducing  quark 
anomalous magnetic moments. 
Introducing quark form factors
results in a faster $Q^2$ fall-off of 
form factors and forces
$m_q(Q^2)$ to drop faster with $Q^2$ to describe
the data. We found that descriptions, that are very
close to those for pointlike quarks and masses 
(\ref{eq:nuc4},\ref{eq:nuc5}), 
can be obtained by introducing quark form factors
\begin{equation}
F_q(Q^2)=1/(1+Q^2/a_q)^2
\label{eq:nuc6}
\end{equation}
with $a_q^{(1)}>18~$GeV$^2$ and $a_q^{(2)} >70~$GeV$^2$
for the wave functions $\Phi_1$ and $\Phi_2$, respectively.
The corresponding quark radii are 
$r^{(1)}_q < r_N/5$ and $r^{(2)}_q < r_N/10$,
where $r_N$ is the mean value of the radii
corresponding to $G_{Ep}(Q^2)$, 
$G_{Mp}(Q^2)$, and $G_{Mn}(Q^2)$.
The $Q^2$-dependencies of quark masses for minimal
values of $a_q$ are 
\begin{eqnarray}
&& m_q^{(1)}(Q^2)=\frac{0.22{\rm GeV}}{1+Q^2/20{\rm GeV}^2},
\label{eq:nuc7}\\
&& m_q^{(2)}(Q^2)=\frac{0.22{\rm GeV}}{1+Q^2/6{\rm GeV}^2}.
\label{eq:nuc8}
\end{eqnarray}
Therefore, in our approach the quark mass
can be in ranges
given by Eqs. (\ref{eq:nuc4},\ref{eq:nuc5})
and (\ref{eq:nuc7},\ref{eq:nuc8}).
As mentioned above,
the results for the nucleon electromagnetic form factors
obtained  taking into account
quark form factors (\ref{eq:nuc6}) and masses
(\ref{eq:nuc7},\ref{eq:nuc8}) are very close to those
for pointlike quarks and masses
(\ref{eq:nuc4},\ref{eq:nuc5}).
For this reason, they are not shown separately in Figs.
\ref{fig:nucleon},\ref{fig:cloud}.

\begin{figure*}[htp]
\begin{center}
\includegraphics[width=12.0cm]{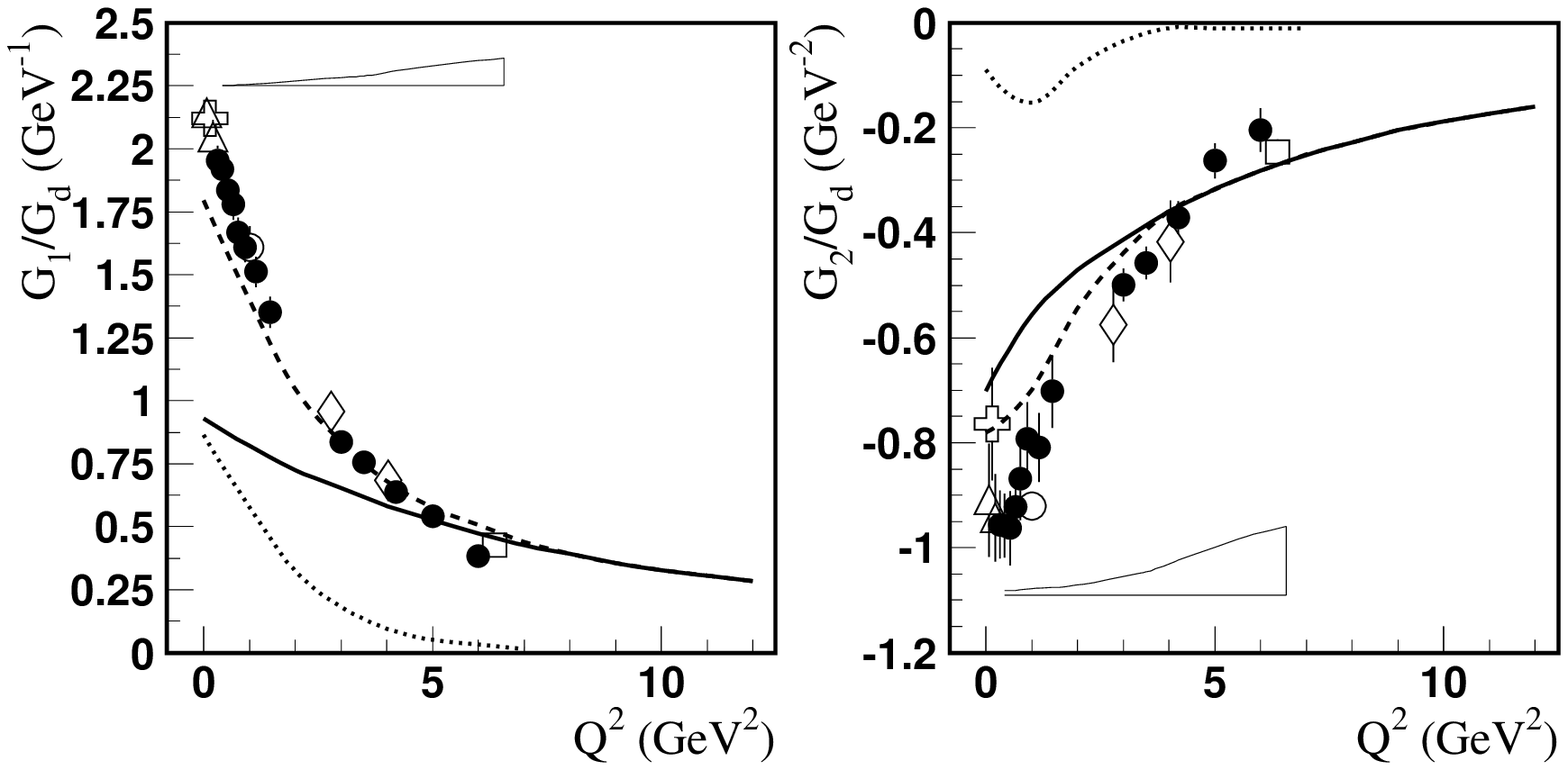}
\vspace{-0.1cm}
\caption{\small
The $\gamma^*p\rightarrow \Delta(1232)$P$_{33}$
transition form factors.
The solid curves correspond to the LF RQM predictions;
the weight factors for the
$3q$ contributions to the $\Delta(1232)$P$_{33}$
are $c_{N^*}^{(1)}\approx 
c_{N^*}^{(2)}=0.53\pm 0.04$ 
for the wave functions 
of Eqs. (\ref{eq:sec9}) and (\ref{eq:sec10}).
The dotted curves correspond to the meson-cloud contributions
obtained in the dynamical model \cite{Sato1}.
The dashed curves present the sum of the $3q$ 
and meson-cloud contributions.
Solid circles are the amplitudes extracted from
the CLAS pion electroproduction data \cite{CLAS}, bands represent model
uncertainties of these results.
The results from other experiments are: open triangles
\cite{Stave2006,Sparveris2007,Stave2008},
open crosses
\cite{Mertz,Kunz,Sparveris2005}, open rhombuses
\cite{Frolov},
open boxes \cite{Vilano},
and open circles
\cite {KELLY1,KELLY2}.
\label{fig:delta}}
\end{center}
\end{figure*}

\begin{figure*}[htp]
\begin{center}
\includegraphics[width=12.0cm]{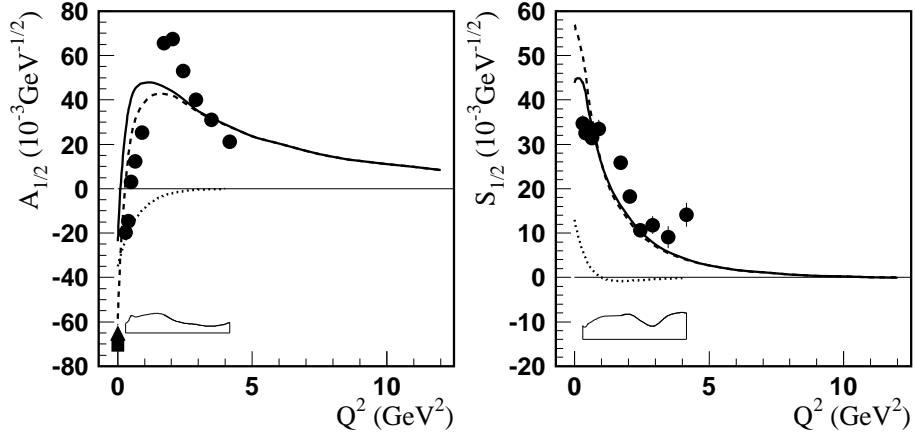}
\vspace{-0.1cm}
\caption{\small
The $\gamma^*p\rightarrow N(1440)$P$_{11}$
transition helicity amplitudes. 
The solid curves are the LF RQM predictions
obtained with
the weight factors 
$c_{N^*}^{(1)}=0.73\pm 0.05$ and
$c_{N^*}^{(2)}=0.77\pm 0.05$
for the $3q$ contribution  
to the $N(1440)$P$_{11}$.
The dotted curves correspond to the $\sigma N$ 
contribution \cite{Obuch}.
The dashed curves present the sum of the $3q$ 
and $\sigma N$ 
contributions.
Solid circles are the amplitudes extracted from
the CLAS pion electroproduction data \cite{CLAS}, bands represent model
uncertainties of these results.
The full box at $Q^2=0$ is
the amplitude extracted from CLAS $\pi$ photoproduction
data \cite{Dugger_pip}.
The full triangle at $Q^2=0$ is
the RPP estimate \cite{PDG}.
\label{fig:p11}}
\end{center}
\end{figure*}

\begin{figure*}[htp]
\begin{center}
\includegraphics[width=12.0cm]{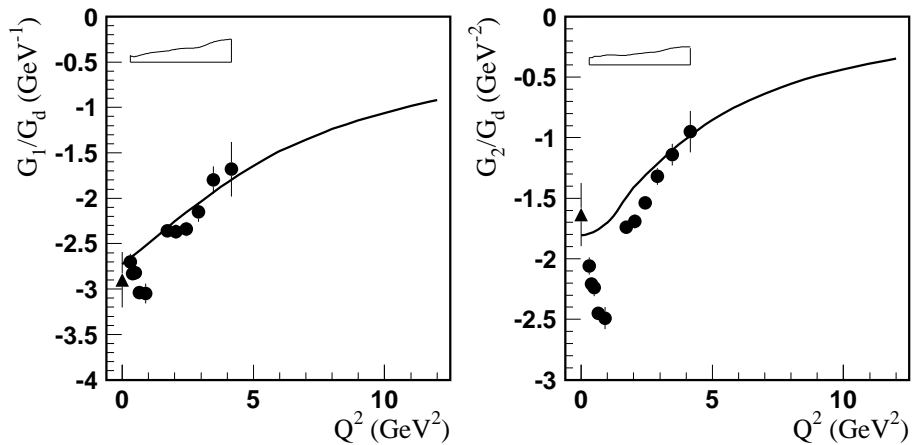}
\vspace{-0.1cm}
\caption{\small
The $\gamma^*p\rightarrow N(1520)$D$_{13}$
transition form factors.
$c_{N^*}^{(1)}=0.78\pm 0.06$,
$c_{N^*}^{(2)}=0.82\pm 0.06$. 
Other legend is as for Fig. \ref{fig:p11}. 
\label{fig:d13}}
\end{center}
\end{figure*}

\begin{figure*}[htp]
\begin{center}
\includegraphics[width=12.0cm]{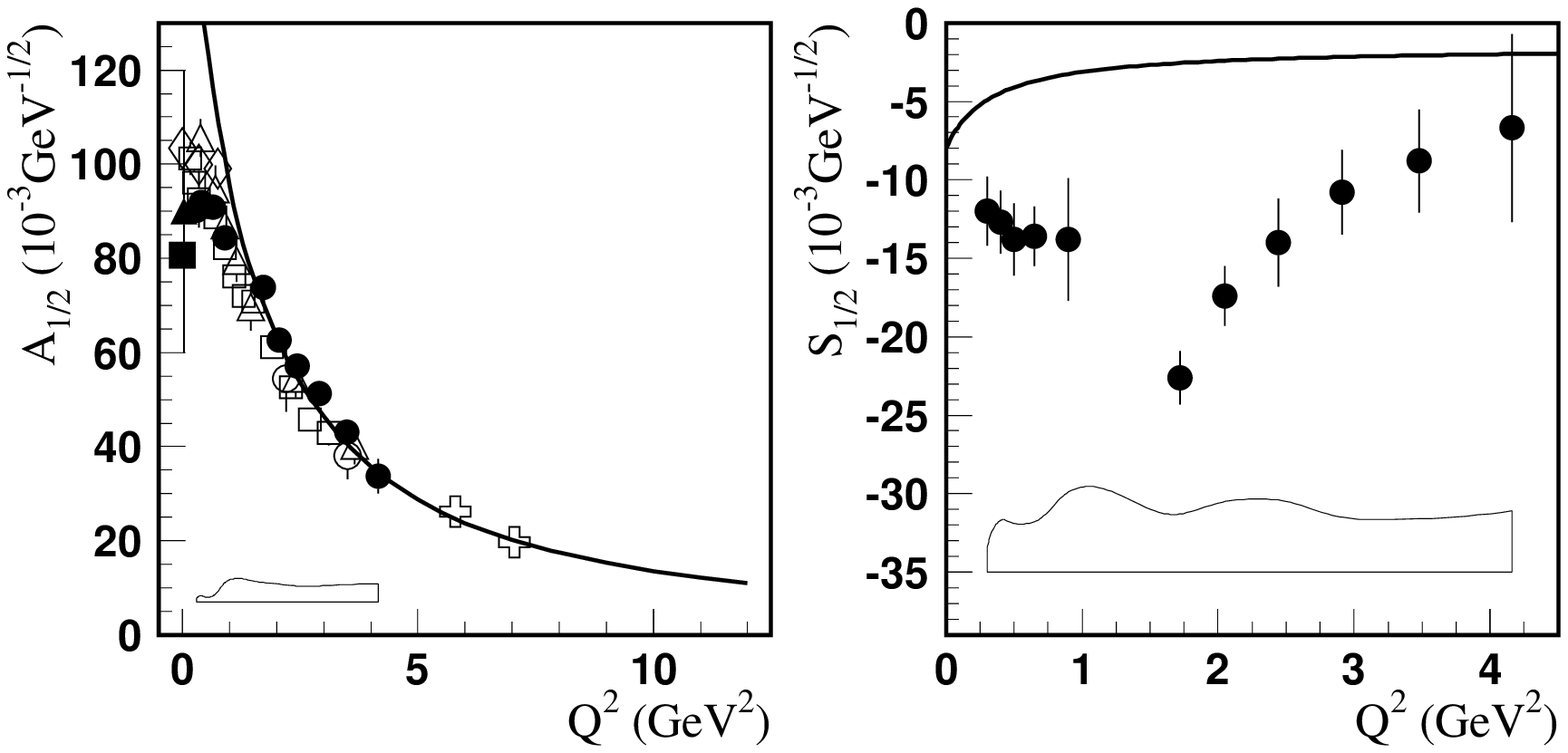}
\vspace{-0.1cm}
\caption{\small
The $\gamma^*p\rightarrow N(1535)$S$_{11}$
transition helicity amplitudes.
The open triangles \cite{Thompson},
open boxes \cite{Denizli}, and open
rhombuses \cite{Aznauryan2005,Aznauryaneta}
are the amplitudes extracted from
the JLab/Hall B $\eta$ electroproduction data;
the open circles \cite{Armstrong}
and open crosses \cite{Dalton}
are the amplitudes extracted from 
the JLab/Hall C $\eta$ electroproduction data.
$c_{N^*}^{(1)}=0.88\pm 0.03$,
$c_{N^*}^{(2)}=0.94\pm 0.03$. 
Other legend is as for Fig. \ref{fig:p11}. 
\label{fig:s11}}
\end{center}
\end{figure*}

\section{Nucleon resonances ${\mathbf \Delta(1232)P_{33}}$, 
${\mathbf N(1440)P_{11}}$, ${\mathbf N(1520)D_{13}}$, 
and ${\mathbf N(1535)S_{11}}$}

No investigations are available that allow for the separation
of the $3q$ and $\pi N$ (or meson-nucleon) contributions
to nucleon resonances. Therefore, 
the weights of the $3q$ contributions to the resonances:
\begin{equation}
|N^*>=c_{N^*}|3q>+...,~~c_{N^*}<1, 
\label{eq:res1}
\end{equation}
are unknown.
We estimate these weights by fitting to 
experimental $\gamma^* N\rightarrow N^*$ amplitudes.
The range of $Q^2$ for the fit has been chosen according to
available information on the possible 
meson-cloud contribution to the transition
amplitudes. 
In Ref. \cite{Sato1}, the dynamical model
has been applied to describe the data on pion
electroproduction on proton in the $\Delta(1232)P_{33}$ 
resonance region at $Q^2\leq 4~$GeV$^2$. As the result,
the contribution that can be
associated with the meson-cloud contribution
to $\gamma^* N\rightarrow \Delta(1232)P_{33}$ 
has been found. Unlike for the nucleon, this contribution
can not be neglected at $Q^2 = 2-4~$GeV$^2$
(see Fig. \ref{fig:delta}).
In Ref. \cite{Sato2}, the coupled-channel
approach has been applied 
to the description of the pion photoproduction data,
and the meson-cloud contribution
to the transverse amplitudes for 
the $N(1440)P_{11}$, $N(1520)D_{13}$,
and $N(1535)S_{11}$ has been found at $Q^2=0$.
The predicted $Q^2$-dependence of this contribution
for absolute values of the amplitudes has been presented.
According to these results, meson-cloud contributions
to $\gamma^* N\rightarrow N(1440)$P$_{11}$,
$N(1520)D_{13}$,
and $N(1535)S_{11}$ 
are negligible at $Q^2 > 2~$GeV$^2$.
Similar results are obtained 
for both transverse and longitudinal 
$\gamma^* N\rightarrow N(1440)$P$_{11}$ 
amplitudes via estimation of
the $\sigma N$ contribution
to this transition \cite{Obuch}
(see Fig. \ref{fig:p11}).

We therefore determine the $3q$ contribution
to the resonances by fitting the experimental
amplitudes 
at $Q^2>4$GeV$^2$ for the
$\Delta(1232)P_{33}$ and
at $Q^2=2.5-4.5$GeV$^2$ for the
$N(1440)P_{11}$, $N(1520)D_{13}$,
and $N(1535)S_{11}$, assuming
that at these $Q^2$ the transition amplitudes
are dominated by the $3q$ contribution. 
The results are shown in Figs. \ref{fig:delta}-\ref{fig:s11}.
For the $\Delta(1232)P_{33}$ and $N(1440)P_{11}$,
we present also the results where the $3q$-core 
is complemented, respectively,  by 
the meson-cloud \cite{Sato1} and $\sigma N$ \cite{Obuch} contributions.
These contributions significantly
improve the agreement with experimental amplitudes
at low $Q^2$.
  
Here we comment on the amplitudes presented
in Figs. \ref{fig:delta}-\ref{fig:s11}.
As it is shown in Refs. \cite{Aznauryan85,Keister},
there are difficulties in the utilization of the LF
approaches \cite{Terentiev,Aznauryan1,Aznauryan2,Capstick1}
for hadrons with spins $J\geq 1$. These difficulties
are not present if Eq. (\ref{eq:sec1}) is used to
calculate only those matrix elements that
correspond to $S'_z=J$ \cite{Aznauryan85}.
This restricts the number of transition form factors
that can be investigated for the resonances  
$\Delta(1232)P_{33}$ and $N(1520)D_{13}$.
As can be seen from Eqs. (\ref{eq:ap11},~\ref{eq:ap12}),  
the matrix elements with $S'_z=\frac{3}{2}$ relate to only two 
transition form factors: $G_1(Q^2)$
and $G_2(Q^2)$. 
Consequently, we can not present the results in terms 
of transition helicity amplitudes for the $\Delta(1232)P_{33}$ 
and $N(1520)D_{13}$, while the resonances with $J=\frac{1}{2}$, i.e. 
$N(1440)P_{11}$ and $N(1535)S_{11}$,
are presented in terms of these  amplitudes.

Using eqs. (\ref{eq:a14} - \ref{eq:a19}),
the transition form factors $G_{1,2}(Q^2)$
for the $\Delta(1232)P_{33}$ and $N(1520)D_{13}$ 
can be related to the  
transition helicity amplitudes. 
For $G_{1}(Q^2)$, the relation has simple form:
\begin{equation}
G_1(Q^2)=\mp \frac{m_{N^*}}{2XQ_{\pm}}
\left(A_{1/2}\pm \frac{1}{\sqrt{3}}A_{3/2}\right),
\label{eq:res2}
\end{equation}
where $Q_{\pm}$ and $X$ are defined
by eqs. (\ref{eq:app14},~\ref{eq:a16})
and the upper and lower symbols correspond,
respectively, to the $\Delta(1232)P_{33}$ and $N(1520)D_{13}$.
For the $\Delta(1232)P_{33}$, it is useful
to also present the following relation:
\begin{equation}
G_1(Q^2)=\sqrt{\frac{3}{2}}\frac{m_{N^*}(m_{N^*}+m_N)}{m_NQ_{+}}
\left(G_M(Q^2)-G_E(Q^2)\right),
\label{eq:res3}
\end{equation}
where $G_M(Q^2)$ and $G_E(Q^2)$ are the form factors
defined in Ref. \cite{Scadron}.

We note that the predictions obtained with different
wave functions (\ref{eq:sec9},\ref{eq:sec10}) and
corresponding masses (\ref{eq:nuc4},\ref{eq:nuc5}),
as well as the predictions found  taking into account
quark form factors (\ref{eq:nuc6}) and masses
(\ref{eq:nuc7},\ref{eq:nuc8}) only differ in the weight factors 
$c_{N^*}$ for the wave functions 
(\ref{eq:sec9}) and 
(\ref{eq:sec10}); 
these factors are given in the figure captions. 
The LF RQM predictions for resonances 
are therefore presented by a single (thick solid) curve.

We also note that the nucleon and $\Delta(1232)P_{33}$ 
as well as the
$N(1440)P_{11}$ are considered as members of the
$[56,0^+]$ and $[56,0^+]_R$ multiplets, respectively.
The $N(1520)D_{13}$ is taken as the state ${}^28_{3/2}$ 
of the multiplet $[70,1^-]$ and
the $N(1535)S_{11}$ as
a mixture of the states ${}^28_{1/2}$ and
${}^48_{1/2}$ in this mutiplet:
\begin{equation}
N(1535)S_{11}=\rm{cos}\theta_S |{}^28_{1/2}>-\rm{sin}\theta_S|{}^48_{1/2}>.
\label{eq:sqtm}
\end{equation}
Here we use the notation
${}^{2S+1}SU(3)_{J}$, which gives the assignment 
according to the $SU(3)$ group, $J$ is
the spin of the resonance, and $S$ is the total spin of quarks.
The mixing angle is taken equal to
$\theta_S=-31^{\circ}$ as found from the hadronic
decays \cite{Isgur_Karl,Hey1975}.
The transition $\gamma^* p \rightarrow {}^48_{1/2}$,
which is forbidden in the single quark transition
model \cite{SQTM}, turned out very small
compared to $\gamma^* p \rightarrow {}^28_{1/2}$
in the LF RQM too.
Therefore, the 
$\gamma^* p \rightarrow N(1535)S_{11}$ amplitudes are
determined mainly by the first term in Eq. (\ref{eq:sqtm}).

\section{Summary}

We have described the nucleon electromagnetic form factors
in a wide range of $Q^2$ by complementing the $3q$-core contribution
with contribution of the pion cloud, and assuming the
constituent quark mass to decrease with increasing $Q^2$.
The pion-cloud contribution is negligible at $Q^2>2~$GeV$^2$,
but it is important to 
describe the neutron electric form factor and 
the dip in the magnetic form factors at very small $Q^2$.
The decreasing quark mass allowed us to compensate the 
rapidly falling form factors with increasing $Q^2$.
The $Q^2$-dependent quark mass is in qualitative agreement 
with results from QCD lattice and Dyson-Schwinger equations.
The mechanism, that generates the running quark mass
within these approaches, can also produce quark form factors
which result in a faster fall-off of the nucleon 
form factors. This, in turn, forces $m_q(Q^2)$ to drop faster with
$Q^2$ in order to describe the data. From the description of the 
nucleon electromagnetic form factors,
we have found empirically the boundaries 
for the quark form factors and the corresponding
boundaries for $m_q(Q^2)$. 

With the LF RQM specified via description of the nucleon
electromagnetic form factors, we have predicted 
the quark core contribution to the electroexcitation
amplitudes of the resonances 
$\Delta(1232)P_{33}$,
$N(1440)P_{11}$, $N(1520)D_{13}$, and  $N(1535)S_{11}$ 
up to $Q^2=12~$GeV$^2$, where   
the weight factor of the $3q$ contribution to the resonance occurs 
as the only parameter.
This parameter was found by fitting to experimental
amplitudes in a  $Q^2$ range, where the meson-cloud contribution 
is expected to be negligible.
The important feature of our predictions is the fact that at these $Q^2$ 
we describe both amplitudes investigated for each 
resonance by fitting a single parameter.

For the $\Delta(1232)P_{33}$ and $N(1440)P_{11}$,
we also present the results where the $3q$-core
is complemented, respectively,  by
the meson-cloud contribution found
in the dynamical model \cite{Sato1} and the $\sigma N$ 
contribution found in Ref. \cite{Obuch} .
These contributions significantly
improve the agreement with experimental amplitudes
at low $Q^2$.

\vspace{0.3cm}

 \section{Acknowledgments}
We acknowledge valuable communications with
T.-S.H. Lee and C.D. Roberts. 
This work was supported by the US Department of Energy under
contract DE-AC05-06OR23177 and the Department of Education
and Science of Republic of Armenia, Grant-11-1C015.

\section{Appendix. 
The relations between the matrix elements (\ref{eq:sec1})
and the $\gamma^* N\rightarrow N(N^*)$ form factors and transition
helicity amplitudes}
\vspace{0.3cm}
\renewcommand\theequation{A\arabic{equation}}
\setcounter{equation} 0

For the nucleon, the matrix elements (\ref{eq:sec1})
are related to the form factors in the
following way:
\begin{eqnarray}
& \frac{1}{2P_z}<N,\frac{1}{2}|J_{em}^{0,3}|N,\frac{1}{2}>|_
{P_z\rightarrow\infty}=F_1,
\label{eq:ap1}
\\
& \frac{1}{2P_z}<N,\frac{1}{2}|J_{em}^{0,3}|N,-\frac{1}{2}>|_
{P_z\rightarrow\infty}=-\frac{Q}{2m_N}F_2,
\label{eq:ap2}
\end{eqnarray}
where $F_1(Q^2)$ and $F_2(Q^2)$ are the Dirac and Pauli
form factors: $F_{1p}(0)=1$, $F_{2N}(0)=\kappa_N$, the 
nucleon anomalous
magnetic moment.
The Sachs form factors are:
\begin{equation}
G_M(Q^2)=F_1+F_2,~~~~
G_E(Q^2)=F_1-\frac{Q^2}{4m_N^2}F_2.
\label{eq:ap4}
\end{equation}

For the resonances with $J^P={\frac{1}{2}}^{\pm}$:

\begin{eqnarray}
&& \frac{1}{2P_z}<N^*,\frac{1}{2}|J_{em}^{0,3}|N,\frac{1}{2}>|_
{P_z\rightarrow\infty}=Q^2G_1,
\label{eq:ap5}
\\
&& \frac{1}{2P_z}<N^*,\frac{1}{2}|J_{em}^{0,3}|N,-\frac{1}{2}>|_
{P_z\rightarrow\infty}\\
&&~~~~~~~~~~~~~~=\frac{\pm
m_{N^*}-m_N}{2}QG_2,
\label{eq:ap6}
\end{eqnarray}
where the upper and lower symbols correspond, respectively,
to $J^P={\frac{1}{2}}^{+}$ and ${\frac{1}{2}}^{-}$
resonances, and 
the form factors are defined by 
\cite{Aznauryan11,Devenish}:
\begin{eqnarray}
&<N^{*}|J_{em}^{\mu}|N>\equiv e\bar{u}(P')
\left(\begin{array}{c}1\\\gamma_5\end{array}\right)
\tilde{J}^{\mu}u(P),
\label{eq:ap7}
\\
&{\tilde{J}}^{\mu} =
\left(k\hspace{-1.8mm}\slash k^{\mu}-k^2\gamma^{\mu}\right)G_1
+\left[k\hspace{-1.8mm}\slash {\cal P}^{\mu}-({\cal P}k)\gamma^{\mu}\right]G_2,
\label{eq:ap8}
\end{eqnarray}
${\cal P}\equiv \frac{1}{2}(P'+P)$, $u(P),u(P')$
are the Dirac spinors.
The relations between
the $\gamma^* N\rightarrow N^{*}$
helicity amplitudes and the form factors $G_1(Q^2),G_2(Q^2)$
are following:
\begin{eqnarray}
&&A_{\frac{1}{2}}=
b\left[2Q^2G_1-(m_{N^*}^2-m_{N}^2)G_2\right],
\label{eq:ap9}
\\
&&S_{\frac{1}{2}}=\pm
b\frac{|{\bf{k}}|}{\sqrt{2}}{\tilde S}_{\frac{1}{2}},
\label{eq:ap10}
\\
&&{\tilde S}_{\frac{1}{2}}=2(m_{N^*}\pm m_N)G_1+
(m_{N^*}\mp m_N)G_2,
\label{eq:app11}
\\
&&b\equiv e\sqrt{\frac{Q_{\mp}}{8m_N(m_{N^*}^2-m_{N}^2)}},
\label{eq:app12}
\\
&& |{\bf{k}}|=\frac{\sqrt{Q_{+}Q_{-}}}{2m_{N^*}},
\label{eq:app13}
\\
&& Q_{\pm}\equiv (m_{N^*}\pm m_N)^2+Q^2.
\label{eq:app14}
\end{eqnarray}

For the resonances with $J^P={\frac{3}{2}}^{\pm}$:

\begin{eqnarray}
&& \frac{1}{2P_z}<N^*,\frac{3}{2}|J_{em}^{0,3}|N,\frac{1}{2}>|_
{P_z\rightarrow\infty}=\nonumber \\
&&-\frac{Q}{\sqrt{2}}
\left[G_1(Q^2)+\frac{\pm m_{N^*}-m_N}{2}G_2(Q^2)\right],
\label{eq:ap11}
\\
&& \frac{1}{2P_z}<N^*,\frac{3}{2}|J_{em}^{0,3}|N,-\frac{1}{2}>|_
{P_z\rightarrow\infty}=
\nonumber
\\
&&~~~~~~~~~~~~~~~~\frac{Q^2}{2\sqrt{2}}G_2(Q^2),
\label{eq:ap12}
\end{eqnarray}
and the form factors are defined by 
\cite{Aznauryan11,Devenish}:
\begin{eqnarray}
&<N^{*}|J_{em}^{\mu}|N>\equiv e\bar{u}
_{\nu}(P')
\left(\begin{array}{c}\gamma_5\\1\end{array}\right)
\Gamma^{\nu\mu}
u(P),
\label{eq:ap13}\\
&\Gamma^{\nu\mu}(Q^2)= G_1{\cal H}_1^{\nu\mu}+
G_2{\cal H}_2^{\nu\mu}+G_3{\cal H}_3^{\nu\mu},
\label{eq:ap14}\\
&{\cal H}_1^{\nu\mu}=k\hspace{-1.8mm}\slash 
g^{\nu\mu}-k^{\nu}\gamma^{\mu}, 
\label{eq:ap15}\\
&{\cal H}_2^{\nu\mu}=k^{\nu}P'^{\mu}-(kP')g^{\nu\mu},
\label{eq:ap16}\\
&{\cal H}_3^{\nu\mu}=k^{\nu}k^{\mu}-k^2g^{\nu\mu},
\label{eq:ap17}
\end{eqnarray}
where $u_{\nu}(P')$ is the generalized Rarita-Schwinger spinor.
The relations between
the $\gamma^* N\rightarrow N^{*}$
helicity amplitudes and form factors for
the $J^P={\frac{3}{2}}^{\pm}$ resonances
are following:

\begin{eqnarray}
&&{A}_{1/2}=h_3X,
~~~{ A}_{3/2}=\pm \sqrt{3}h_2X,
\label{eq:a14}
\\
&&{S}_{1/2}=h_1\frac{|\bf{k}|}{\sqrt{2}m_{N^*}}X,
\label{eq:a15}
\\
&&X\equiv e\sqrt{\frac{Q_{\mp}}
{48m_N(m_{N^*}^2-m_{N}^2)}},
\label{eq:a16}
\end{eqnarray}
where
\begin{eqnarray}
&&h_1(Q^2)=\pm 4m_{N^*}G_1(Q^2)+4m_{N^*}^2G_2(Q^2)+\nonumber \\
&&~~~~~~~~~~~2(m_{N^*}^2-m_N^2-Q^2)G_3(Q^2),
\label{eq:a17}\\
&&h_2(Q^2)=-2(\pm m_{N^*}+ m_N)G_1(Q^2)-\nonumber \\
&&(m_{N^*}^2-m_N^2-Q^2)G_2(Q^2)+
2Q^2G_3(Q^2),
\label{eq:a18}\\
&&h_3(Q^2)=\mp\frac{2}{m_{N^*}}[Q^2+m_N(\pm
m_{N^*}+m_N)]G_1(Q^2)+\nonumber \\
&&(m_{N^*}^2-m_N^2-Q^2)G_2(Q^2)-2Q^2G_3(Q^2).
\label{eq:a19}
\end{eqnarray}

\end{document}